\documentclass[twocolumn,showpacs,preprintnumbers]{revtex4}
\usepackage{graphics}
\usepackage{graphicx}

\begin{document}

\title{A spin chain with spiral orders: perspectives of quantum information
and mechanical response}
\author{Shi-Jian Gu}
\altaffiliation{Email: sjgu@phy.cuhk.edu.hk}
\author{Wing-Chi Yu}
\author{Hai-Qing Lin}
\date{\today }

\begin{abstract}
In this paper, we study the ground state of a one-dimensional exactly
solvable model with a spiral order. While the model's energy spectra is the
same as the one-dimensional transverse field Ising model, its ground state
manifests spiral order with various periods. The quantum phase transition
from a spiral-order phase to a paramagnetic phase is investigated in
perspectives of quantum information science and mechanics. We show that the
modes of the ground-state fidelity and its susceptibility can tell the
change of periodicity around the critical point. We study also the spin
torsion modulus which defines the coefficient of the potential energy stored
under a small rotation. We find that at the critical point, it is a
constant; while away from the critical point, the spin torsion modulus tends
to zero.
\end{abstract}

\maketitle


\address{Department of Physics and ITP, The Chinese University of Hong Kong,
Hong Kong, China}

\address{Department of Physics and ITP, The Chinese University of Hong Kong,
Hong Kong, China}

\address{Department of Physics and ITP, The Chinese University of Hong Kong,
Hong Kong, China}




\section{Introduction}

Spiral (or helical) order is a very important phenomenon in condensed matter
physics, especially in soft matters. A typical case is the well-known
deoxyribonucleic acid DNA. In solid state, the spiral order can also be
found in spin systems with competing exchange interactions. Studies on the
spiral order in these systems could be traced back to the later of 1950's%
\cite{Yoshimori,Kaplan}. Up to now, many works on the spiral order have been
published\cite{Freiser,Redner,Rosenfeld,Kaplan2009,JCao2003}. Among these
studies, an interesting example is the one-dimensional XXZ spin chain with
unparallel boundary fields \cite{JCao2003}. The model's ground state shows a
spiral order. Moreover, it has been solved exactly using the Bethe-ansatz
method \cite{JCao2003}, hence many physical properties can be well studied
without any approximation.

However, as far as we know, no one addressed the critical phenomena
occurring in spin systems with spiral order in perspective of quantum
information science \cite{Nielsen1}. So the main motivation of the present
paper is to study the spiral order transition in terms of the concept of
fidelity from the quantum information science. For this purpose, we consider
a quantum spin chain whose ground state consists of a spiral order phase and
a paramagnetic phase. The quantum phase transition occurring between the two
phases is studied in perspectives of quantum information science and spin
torsion modulus. We show that the modes of the ground-state fidelity and its
susceptibility can tell the change of the period of the spiral order around
the critical point. We study also the spin torsion modulus, which defines
the coefficient of the potential energy stored under a small rotation,
around the critical point.

The paper is organized as follows: In section \ref{sec:model}, we introduce
the model and solve it exactly in a standard procedure. The quantum phase
transitions from a spiral order phase to a paramagnetic phase is also
discussed. In section \ref{sec:fs}, we study the quantum phase transition in
term of mode fidelity and its susceptibility. We show that the mode fidelity
and its susceptibility can not only detect the critical point, but also
indicate which mode undergoes a significant change at the critical point. In
section \ref{sec:spinmodulus}, we study the spin torsion modulus of the
model at the zero temperature. We find away from the critical point, the
normalized spin torsion modulus tends to zero as the system size increases;
while at the critical point, it is a constant. Finally, a summary is given
in section \ref{sec:sum}.

\section{The model Hamiltonian and its exact solution}

\label{sec:model}

We consider a spin chain described by the Hamiltonian
\begin{equation}
H=-\sum_{j=1}^{N}\left( s_{j}^{x}s_{j+1}^{x}+hs_{j}^{z}\right) ,
\label{eq:Hamiltonian1}
\end{equation}%
where $s_{j}$ is the twisted 1/2 spin operators. They are defined as
\begin{eqnarray}
s_{j}^{x} &\equiv &\exp \left[ \frac{i}{2}j\phi \sigma _{j}^{z}\right]
\sigma _{j}^{x}\exp \left[ -\frac{i}{2}j\phi \sigma _{j}^{z}\right] ,
\nonumber \\
s_{j}^{y} &\equiv &\exp \left[ \frac{i}{2}j\phi \sigma _{j}^{z}\right]
\sigma _{j}^{y}\exp \left[ -\frac{i}{2}j\phi \sigma _{j}^{z}\right] ,
\nonumber \\
s_{j}^{z} &\equiv &\sigma _{j}^{z},
\end{eqnarray}%
where $\phi $ is an angle and $\sigma =(\sigma ^{x},\sigma ^{y},\sigma ^{z})$
are the Pauli matrices. It is easy to check that $s=(s^{x},s^{y},s^{z})$
satisfies the su(2) Lie algebra, i.e.
\begin{eqnarray}
\left[ s_{j}^{x},s_{j}^{y}\right] &=&2is_{j}^{z},  \nonumber \\
\left[ s_{j}^{y},s_{j}^{z}\right] &=&2is_{j}^{x},  \nonumber \\
\left[ s_{j}^{z},s_{j}^{x}\right] &=&2is_{j}^{y}.
\end{eqnarray}%
If we introduce
\begin{equation}
\sigma ^{+}=\frac{1}{2}\left( \sigma ^{x}+i\sigma ^{y}\right) ,\text{ }%
\sigma ^{-}=\frac{1}{2}\left( \sigma ^{x}-i\sigma ^{y}\right) ,
\end{equation}%
then the Hamiltonian becomes%
\begin{eqnarray}
H &=&-\sum_{j=1}^{N}\left[ \left( e^{-i\phi }\sigma _{j}^{+}\sigma
_{j+1}^{-}+h.c.\right) \right.  \nonumber \\
&&\left. +\left( e^{i(2j+1)\phi }\sigma _{j}^{+}\sigma
_{j+1}^{+}+h.c.\right) \right] -h\sum_{j=1}^{N}\sigma _{j}^{z}.
\end{eqnarray}%
The Hamiltonian either exchanges the state of a pair of anti-parallel spins,
or flip two upward spins to downward or vice versa., so if we define a
parity operator%
\begin{equation}
P=\prod_{j=1}^{N}\sigma _{j}^{z},
\end{equation}%
and the Hamiltonian cannot change the parity of the state, that is.
\begin{equation}
\lbrack H,P]=0.
\end{equation}%
Therefore, we have two subspaces corresponding to parity $P=\pm 1$
respectively.

If $\phi =0$, the Hamiltonian describes the one-dimensional transverse field
Ising model \cite{Sachdev,PPfeuty70,RJElliott70} with ferromagnetic order%
\begin{equation}
H_{\text{Ising}}=-\sum_{j=1}^{N}\left( \sigma _{j}^{x}\sigma
_{j+1}^{x}+h\sigma _{j}^{z}\right) .
\end{equation}%
While if $\phi =\pi $, it describes the same model but with
antiferromagnetic order. Mathematically, if we introduce a unitary
transformation%
\begin{equation}
R(\phi )=\prod\limits_{j=1}^{N}R_{j}(\phi )=\exp \left[ \frac{i\phi }{2}%
\sum\limits_{j=1}^{N}j\sigma _{j}^{z}\right] ,
\end{equation}%
The Hamiltonian (\ref{eq:Hamiltonian1}) can be transformed to the
one-dimensional transverse field Ising model, i.e.
\begin{equation}
H_{\text{Ising}}=R^{\ast }(\phi )HR(\phi ).
\end{equation}%
Therefore, the energy spectra of the Hamiltonian (\ref{eq:Hamiltonian1}) is
the same as the quantum Ising model if the corresponding boundary conditions
are satisfied.

Nevertheless, what we are interested in this paper is the change in the
structure of the ground-state wavefunction around the quantum critical
point. For this purpose, we still want to take the standard procedure to
diagonalize the Hamiltonian (\ref{eq:Hamiltonian1}) and get the ground-state
wavefunction explicitly. The procedure consists of three transformations, as
shown below.

\begin{figure}[tbp]
\includegraphics[width=8cm]{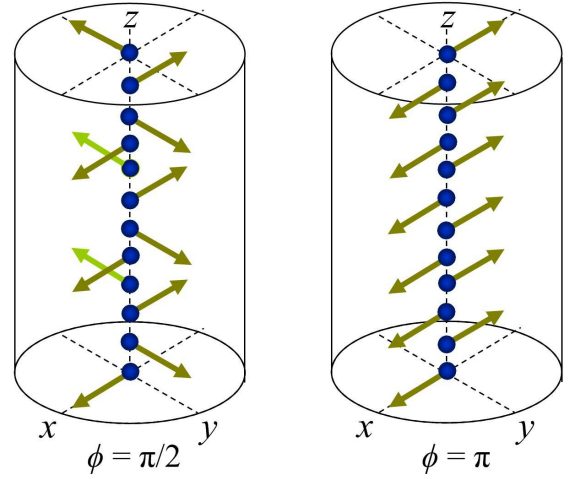}
\caption{(color online) A sketch of the spiral order. LEFT: a spiral with $%
\protect\phi=\protect\pi/2$ and period $S=4$. RIGHT: a spiral order with $%
\protect\phi=\protect\pi$ and period $S=2$, which actually corresponds to an
antiferromagnetic order.}
\label{fig:spiralorder}
\end{figure}

\textit{The Jordan-Wigner transformation: }The Jordan-Wigner transformation
maps 1/2 spins to spinless fermions, that is%
\begin{eqnarray}
\sigma _{n}^{+} &=&\exp \left[ i\pi \sum_{j=1}^{n-1}c_{j}^{\dagger }c_{j}%
\right] c_{n}=\prod_{j=1}^{n-1}\sigma _{j}^{z}c_{n},  \nonumber \\
\sigma _{n}^{-} &=&\exp \left[ -i\pi \sum_{j=1}^{n-1}c_{j}^{\dagger }c_{j}%
\right] c_{n}^{\dagger }=\prod_{j=1}^{n-1}\sigma _{j}^{z}c_{n}^{\dagger },
\nonumber \\
\sigma _{n}^{z} &=&1-2c_{n}^{\dagger }c_{n}.
\end{eqnarray}%
Then the Hamiltonian becomes%
\begin{eqnarray}
H &=&-\sum_{j=1}^{N}\left[ \left( e^{i\phi }c_{j}^{\dagger
}c_{j+1}+h.c.\right) \right.  \label{eq:Hamiltonianfermi} \\
&&\left. +\left( e^{-i(2j+1)\phi }c_{j}^{\dagger }c_{j+1}^{\dagger
}+h.c.\right) \right] -\sum_{j=1}^{N}h\left( 1-2c_{j}^{\dagger }c_{j}\right)
.  \nonumber
\end{eqnarray}%
If the twist angle satisfies $\phi =2n\pi /N,n=0,1,...,N-1$, the fermionic
version of the Hamiltonian has antiperiodic boundary conditions for $P=1$
and periodic boundary conditions for $P=-1$, respectively. In this case, the
energy spectra of the Hamiltonian is the same as the quantum Ising model.

\textit{The Fourier transformations:} Clearly, if $h$ is infinite, all spins
are aligned to the $z$ direction, then $P=1$. So the ground state locates in
the subspace of $P=1$. Under this condition, we can make Fourier
transformations
\begin{eqnarray}
c_{j} &=&\frac{1}{\sqrt{N}}\sum_{k}e^{-ikj}c_{k},\text{ \ \ }  \nonumber \\
c_{j}^{\dagger } &=&\frac{1}{\sqrt{N}}\sum_{k}e^{ikj}c_{k}^{\dagger },
\end{eqnarray}%
where the momentum $k$s are chosen as%
\begin{equation}
k=\frac{(2n+1)\pi }{N}.
\end{equation}%
The Hamiltonian then becomes%
\begin{eqnarray}
H &=&-\sum_{k}\left[ \left( 2\cos (k-\phi )-2h\right) c_{k}^{\dagger
}c_{k}\right.  \label{eq:Hamiltonian_Isingk} \\
&&\left. -i\sin (k-\phi )\left( c_{k}^{\dagger }c_{-k+2\phi }^{\dagger
}+c_{k}c_{-k+2\phi }\right) +h\right]  \nonumber
\end{eqnarray}

\textit{The Bogoliubov transformation:} The above quadratic Hamiltonian can
be further diagonalized under the Bogoliubov transformation:

\begin{eqnarray}
c_{k} &=&u_{k}b_{k}+iv_{k}b_{-k+2\phi }^{\dagger },  \nonumber \\
c_{k}^{\dagger } &=&u_{k}b_{k}^{\dagger }-iv_{k}b_{-k+2\phi },  \nonumber \\
c_{-k+2\phi } &=&u_{-k+2\phi }b_{-k+2\phi }+iv_{-k+2\phi }b_{k}^{\dagger },
\nonumber \\
c_{-k+2\phi }^{\dagger } &=&u_{-k+2\phi }b_{-k+2\phi }^{\dagger
}-iv_{-k+2\phi }b_{k},  \label{eq:Bogtran}
\end{eqnarray}%
where $b_{k}$ and $b_{k}^{\dagger }$ are also fermionic operator and satisfy
the same anticommutation relation as $c_{k}$ and $c_{k}^{\dagger }$. Because
of this, one can find the coefficients in the transformation (\ref%
{eq:Bogtran}) should satisfy the following condition

\begin{eqnarray}
u_{k} &=&u_{-k+2\phi },\text{ \ }  \nonumber \\
v_{k} &=&-v_{-k+2\phi },\text{ \ \ }u_{k}^{2}+v_{k}^{2}=1.
\end{eqnarray}%
So we can introduce trigonal relation
\begin{equation}
v_{k}=\sin \theta _{k},\text{\ }u_{k}=\cos \theta _{k}.
\end{equation}%
Inserting the Bogoliubov transformation into Eq. (\ref{eq:Hamiltonian_Isingk}%
), the Hamiltonian can be diagonalized under the conditions\
\begin{eqnarray}
\cos 2\theta _{k} &=&\frac{h-\cos (k-\phi )}{\sqrt{1-2h\cos (k-\phi )+h^{2}}}%
,  \nonumber \\
\sin 2\theta _{k} &=&\frac{-\sin (k-\phi )}{\sqrt{1-2h\cos (k-\phi )+h^{2}}}.
\end{eqnarray}%
Then the Hamiltonian becomes

\begin{equation}
H=\sum_{k}\epsilon (k)\left( 2b_{k}^{\dagger }b_{k}-1\right) ,
\label{eq:DiagonalizedH}
\end{equation}%
where $\epsilon (k)$ is the dispersion relation
\begin{equation}
\epsilon (k)=\sqrt{1-2h\cos (k-\phi )+h^{2}}.  \label{eq:dispersion}
\end{equation}%
From Eq. (\ref{eq:DiagonalizedH}), we can judge that the ground state is a
vacuum state of $b_{k}$. Then the ground-state energy can be calculated as%
\begin{equation}
E_{0}(\phi )=-\sum_{k}\sqrt{1-2h\cos (k-\phi )+h^{2}}.  \label{eq:GSenergy}
\end{equation}

From the dispersion of quasi particles [Eq. (\ref{eq:dispersion})], we can
see that the system is gapless only at $h_{c}=1$ at which a quantum phase
transition is expected to occur. If $h\neq 1$, the system is gapped. The two
phases can be understood from the corresponding limiting cases. If $h\gg 1$,
all spin are oriented along the $z$-direction. It is a paramagnetic phase.
If $h\sim 0$, the spin chain shows a spiral order. The period of the spiral
order is determined by $\phi $. In Fig. 1, we show a sketch of spiral orders
with $\phi =\pi /2$ and $\phi =\pi \ $respectively. The $\phi =\pi $ mode
actually corresponds to the antiferromagnetic order. The competition between
spiral orders and paramagnetic state leads to the quantum phase transitions
at $h_{c}=1$.

Since the ground state $\left\vert \Psi _{0}\right\rangle $ must satisfy $%
b_{k}\left\vert \Psi _{0}\right\rangle =0$, to obtain the ground state, we
need to express the operator $b_{k}(b_{-k+2\phi })$ in terms of $%
c_{k}(c_{k}^{\dagger })$,
\begin{eqnarray}
b_{k} &=&\cos \theta _{k}c_{k}-i\sin \theta _{k}c_{-k+2\phi }^{\dagger },
\nonumber \\
b_{-k+2\phi } &=&\cos \theta _{k}c_{-k+2\phi }+i\sin \theta
_{k}c_{k}^{\dagger }.
\end{eqnarray}%
The condition $b_{k}\left\vert \Psi _{0}\right\rangle =0$ gives
\begin{equation}
a=\cos \theta _{k},b=0,c=0,d=i\sin \theta _{k}.
\end{equation}%
The ground state takes the form%
\begin{equation}
|\Psi _{0}(h)\rangle =\prod_{k\text{ pairs}}\left( \cos \theta _{k}|0\rangle
_{k}|0\rangle _{-k+2\phi }+i\sin \theta _{k}|1\rangle _{k}|1\rangle
_{-k+2\phi }\right) .  \label{eq:gswavefunctionIsing}
\end{equation}

Since the parity here is even, the excited states can be obtained by
applying $2n$ $b_{k}^{\dagger }$s with different $k$s to the ground state.
The number of states which belong to this subspace is not $2^{N}$, but $%
\sum_{2j}^{N}=2^{N-1}$. With a similar procedure, we can obtained another $%
2^{N-1}$ in the subspace of $P=-1$. Then the total number of state is $2^{N}$
and the Hilberb space is complete.

\section{Mode fidelity, its susceptibility, and spiral order transitions}

\label{sec:fs}

\begin{figure}[tbp]
\includegraphics[width=10cm]{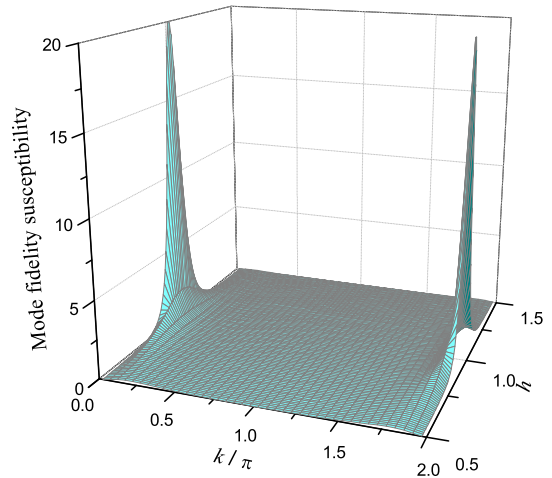}
\caption{(color online) 3D surface figure for the mode fidelity
susceptibility as a function of $k$ and $h$. Here $\protect\phi=0, N=40$.}
\label{fig:modefs_L40_phi000PI}
\end{figure}

\begin{figure}[tbp]
\includegraphics[width=10cm]{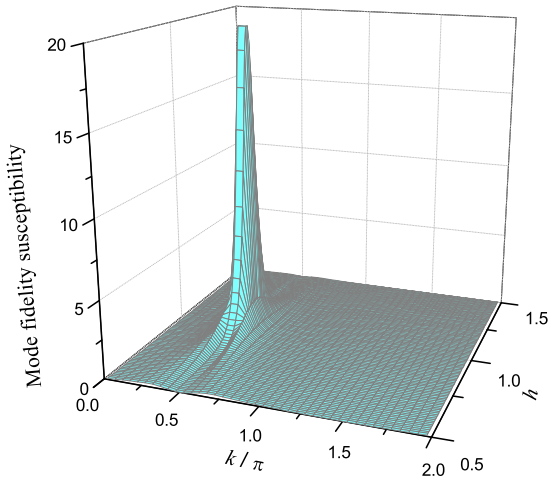}
\caption{(color online) 3D surface figure for the mode fidelity
susceptibility as a function of $k$ and $h$. Here $\protect\phi=\protect\pi%
/2, N=40 $.}
\label{fig:modefs_L40_phi050PI}
\end{figure}

In this section, we are going to study the quantum phase transition
occurring at the critical point $h_{c}=1$ in terms of a concept called
fidelity emerging from quantum information science. The fidelity measures
the similarity between two states. It has been used to study quantum phase
transitions in recent years because it can describe the structure change of
the ground-state wavefunction as the system varies across a quantum critical
point \cite%
{Pzanardi2006,HQZhou0701,WLYou07,SChen07,Venuti07,SJGU08,HQZhou2008,SJGUreview}%
. According to previous publication record, it is commonly believed that the
fidelity can not only detect conventional phase transitions described by
Landau's symmetry-breaking theorem, but also some unconventional phase
transitions, such as topological quantum phase transitions and
Kosterlitz-Thouless phase transitions.

For the present model, the fidelity between $h$ and $h^{\prime }$ can be
calculated as
\begin{equation}
F(h,h^{\prime })=|\langle \Psi _{0}(h^{\prime })|\Psi _{0}(h)\rangle
|=\prod_{k\text{ pairs}}\cos (\theta _{k}-\theta _{k}^{\prime }).
\label{eq:Ising_fidelityexpression}
\end{equation}%
If $\delta h=h-h^{\prime }$ is small, $F(h,h^{\prime })$ shows a minimum
around the critical point due to the big difference between two states
coming from the two phases. However, $F(h,h^{\prime })$ still depends on $%
\delta h$, which is rather artificial. In order to see the leading
contribution to the fidelity, people usually expand the fidelity in terms of
$\delta h$, i.e.
\begin{equation}
F(h,h^{\prime })\simeq 1-\frac{\left( \delta h\right) ^{2}}{2}\chi _{F}(h),
\end{equation}%
where $\chi _{F}(h)$ is the fidelity susceptibility. From Eq. (\ref%
{eq:Ising_fidelityexpression}), the fidelity susceptibility can be
calculated as
\begin{eqnarray}
\chi _{F}(h) &=&\sum_{k\text{ pairs}}\left( \frac{d\theta _{k}}{dh}\right)
^{2} \\
&=&\frac{1}{8}\sum_{k}\frac{\sin ^{2}(k-\phi )}{[1-2h\cos (k-\phi
)+h^{2}]^{2}}.
\end{eqnarray}%
Clearly, the fidelity susceptibility takes the same form as that of the
one-dimensional transverse field Ising model except a phase shift in $k$
space. The phase shift does not affect the value of the fidelity
susceptibility. So around the critical point, the fidelity susceptibility
has the same singular behavior as that in the Ising model \cite{Pzanardi2006}%
, i.e.%
\begin{equation}
\frac{\chi _{F}(h)}{N}\sim \frac{1}{|h-h_{c}|}.
\end{equation}%
If we introduce $\alpha $, $\mu $, $d$, $\nu $ as the critical exponent of $%
\chi _{F}(h)$, scaling dimension of $\chi _{F}(h)$ at the critical point,
scaling dimension of $\chi _{F}(h)$ away from the critical point, and the
critical exponent of the correlation length, they satisfy a scaling relation
$\alpha =(\mu -d)\nu $ \cite{Venuti07,SJGU08}. So the fidelity
susceptibility can not only witness the quantum phase transition, but also
the universality class of the transition. Except for these functions,
however, it seems that the fidelity susceptibility can not tell more about
the quantum phase transition. In order to see more details occurring around
the critical point, we decompose the fidelity susceptibility and introduce
the mode fidelity susceptibility $\chi _{F}(h,k)$, which defines the
changing rate of a specific mode as the driving parameter $h$ varies. For
the present model, the mode fidelity susceptibility takes the form
\begin{equation}
\chi _{F}(h,k)=\frac{1}{8}\frac{\sin ^{2}(k-\phi )}{[1-2h\cos (k-\phi
)+h^{2}]^{2}}.
\end{equation}

\begin{figure}[tbp]
\includegraphics[width=10cm]{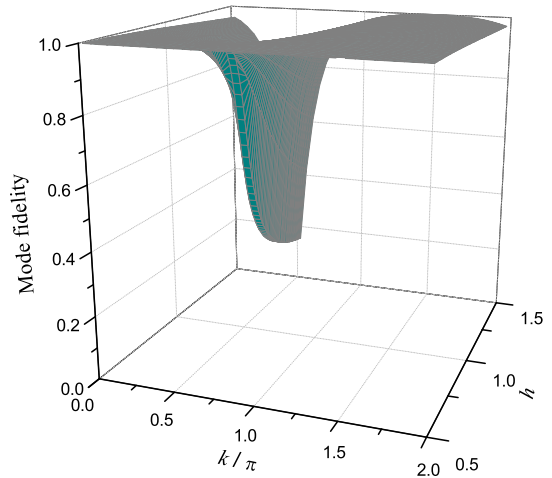}
\caption{(color online) 3D surface figure for the mode fidelity as a
function of $k$ and $h$. Here $\protect\phi=\protect\pi/2, N=40 $, and
starting $h=0.5$.}
\label{fig:modef_L40_phi050PI}
\end{figure}

\begin{figure}[tbp]
\caption{(color online) 3D surface figure for the first derivative of mode
fidelity as a function of $k$ and $h$. Here $\protect\phi=\protect\pi/2,
N=40 $, and starting $h=0.5$.}
\label{fig:devmodef_L40_phi050PI}\includegraphics[width=10cm]{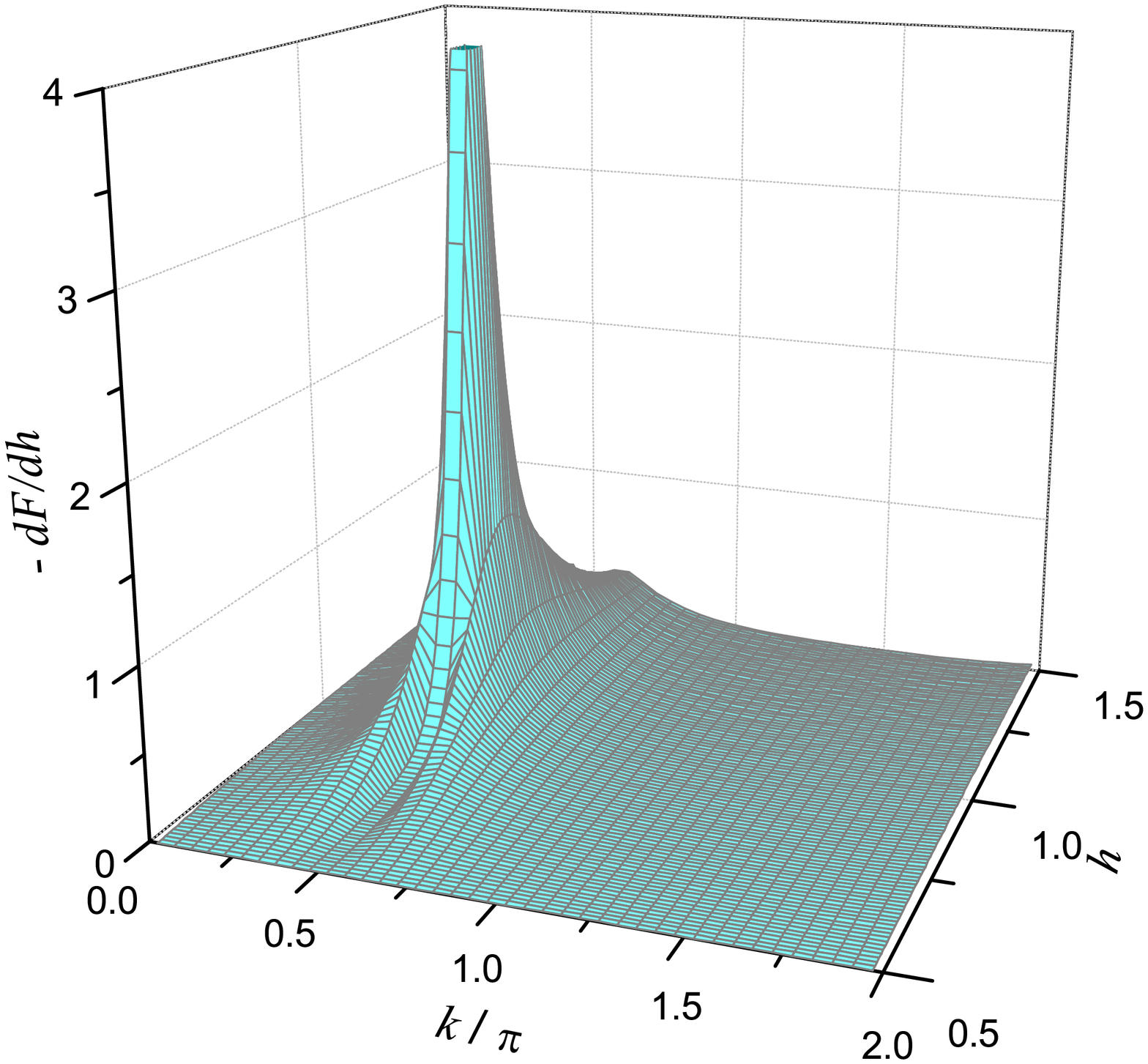}
\end{figure}

\begin{table}[tbp]
\caption{The critical exponent $\protect\alpha$, the scaling dimension at
the critical point $\protect\mu$, and the critical exponent of the
correlation length for the fidelity susceptibility, the relevant mode
fidelity susceptibility, the fidelity per site, and the mode fidelity.}
\label{tab:critcalexp}
\begin{center}
\begin{ruledtabular}
\begin{tabular}{c c c c c}
 & $\alpha$ & $\mu$ & $\nu$ & $d$   \\
\hline fidelity susceptibility  & 1 & 2  & 1 & 1
\\
\hline mode fidelity susceptibility ($k=\phi$) & 2 & 2 & 1 & 0
  \\
\hline fidelity per site & 0 & 0  & 1 & 1
  \\
\hline mode fidelity ($k=\phi$) & 1 & 1  & 1 & 0
  \\
\end{tabular}
\end{ruledtabular}
\end{center}
\end{table}

To see the validity of the mode fidelity susceptibility, we firstly let $%
\phi =0$, which corresponds to the one-dimensional transverse field Ising
model with ferromagnetic order. Around the critical point, $h_{c}=1$, a
quantum phase transition from a ferromagnetic phase to a paramagnetic phase
occurs. Since the ferromagnetic order refers to $k=0$ mode, we expect the $%
k=0$ mode in the ground-state wavefunction undergoes a significant change
around the critical point, while other irrelevant modes do not show
significant changes. We take a sample of $N=40$ sites as an example, and
show the above interpretation in Fig. \ref{fig:modefs_L40_phi000PI}.

Generally, let $\phi =2n\pi /N,$which corresponds to a spiral order with
period $S=N/n$, and $k=(2m+1)\pi /N$, then $k-\phi =[2(m-n)+1]\pi /N$. If $%
m=n$, we can obtain
\begin{equation}
\chi _{F}(h=h_{c},k)\sim \frac{N^{2}}{8\pi ^{2}}.
\end{equation}%
Actually, iff $k-\phi \sim 1/N$, $\chi _{F}(h,k)$ diverges in order of $%
N^{2} $ at the critical point. On the other hand, if we let $k=\phi $
directly, we find%
\begin{equation}
\chi _{F}(h,k)\sim \frac{1}{16}\frac{1}{|h-h_{c}|^{2}},
\end{equation}%
around the critical point. Therefore, for the mode fidelity susceptibility, $%
\alpha =2$, $\mu =2$, $d=0$, $\nu =1$. The scaling relation $\alpha =\nu
(\mu -d)$ is satisfied. Clearly $\chi _{F}(h,k)$ diverges more quickly than
the normalized fidelity susceptibility $\chi _{F}(h)/N$ whose critical
exponent is 1. On the other hand, if $k-\phi $ is not proportional to$\
O(1/N)$, $\chi _{F}(h,k)$ is intensive and independent of the system size $N$%
. So $\chi _{F}(h,k)$ shows no singular behavior around the critical point.

As a numerical demonstration, we take a sample of $N=40$ sites and $\phi
=\pi /2$ as an example. We show the mode fidelity susceptibility as a
function of $\phi $ and $h$ in Fig. \ref{fig:modefs_L40_phi050PI}. From the
figure, we can see that the mode fidelity susceptibility shows singular
behavior at the mode $\phi =\pi /2$, and no singular behavior elsewhere.
From this point of view, the mode fidelity susceptibility might be more
efficient to describe the quantum phase transition occurring in the present
model.

Now we compare the mode fidelity with another related and well studied
concept, i.e., the fidelity per site \cite{HQZhou2008}. For the present
model, the fidelity per site can be calculated as%
\begin{eqnarray}
\ln \mathcal{F}(h,h^{\prime }) &=&\lim_{N\longrightarrow \infty }\frac{1}{N}%
\sum_{k\text{ pairs}}\ln [\cos (\theta _{k}-\theta _{k}^{\prime })], \\
&=&\frac{1}{4\pi }\int_{0}^{2\pi }\ln [\cos (\theta _{k}-\theta _{k}^{\prime
})]dk.
\end{eqnarray}%
It is not difficult to show that the first derivative of the fidelity per
site \cite{HQZhou2008}%
\begin{equation}
\frac{d\ln \mathcal{F}(h,h^{\prime })}{dh}\sim \ln |h-h_{c}|,
\end{equation}%
around the critical point. Therefore, for the fidelity per site, $\alpha =0$.

However, if we focus on the mode of fidelity,
\begin{equation}
F(h,h^{\prime },k)=\cos (\theta _{k}-\theta _{k}^{\prime }),
\end{equation}%
it can show more detail about the change in each mode. We show $%
F(h,h^{\prime },k)$ as a function of $h^{\prime }$ and $k$ in Fig. \ref%
{fig:modef_L40_phi050PI} for the case of $N=40$ and $\phi =\pi /2$. From the
figure, we can see the $k=\pi /2$ decreases the most quickly around the
critical point. To see the changing rate of the mode fidelity, we can take
its first derivative%
\begin{equation}
\frac{dF(h,h^{\prime },k)}{dh^{\prime }}=-\frac{\sin (\theta _{k}-\theta
_{k}^{\prime })}{2}\frac{|\sin (k-\phi )|}{1-2h\cos (k-\phi )+h^{2}}.
\end{equation}%
Fig. \ref{fig:devmodef_L40_phi050PI} shows a 3D map of $dF(h,h^{\prime
},k)/dh^{\prime }$ as a function of $h^{\prime }$ and $k$. Around the
critical point, the most relevant mode scales like
\begin{equation}
\left. \frac{dF(h,h^{\prime },k)}{dh^{\prime }}\right\vert _{k\simeq \phi
}\sim \frac{1}{|h-h_{c}|},
\end{equation}%
\bigskip and
\begin{equation}
\left. \frac{dF(h,h^{\prime },k)}{dh^{\prime }}\right\vert _{k\simeq \phi
,h=1}\sim N.
\end{equation}%
Therefore, for the mode fidelity at $k=\phi $, $\alpha =0,\mu =1,\upsilon
=1,d=0$, the scaling relation is also satisfied. While if $k\neq \phi $, $%
dF(h,h^{\prime },k)/dh\ $is trivial around the critical point. That is the
mode fidelity can not only tell us which mode of the wavefunction undergoes
a significant change around the critical point, but might be more singular
than the fidelity per site.

As a brief summary, we list all the relevant exponents in Table \ref%
{tab:critcalexp} for comparisons. From the table, we can see that the mode
fidelity and its susceptibility are much singular than the fidelity per site
and normalized fidelity susceptibility. Furthermore, they can tell us which
mode undergoes a significant change around the critical point. Therefore,
they are more efficient to describe the critical phenomena occurring in this
model.

\section{Spin torsion modulus}

\label{sec:spinmodulus}

\begin{figure}[tbp]
\includegraphics[width=9cm]{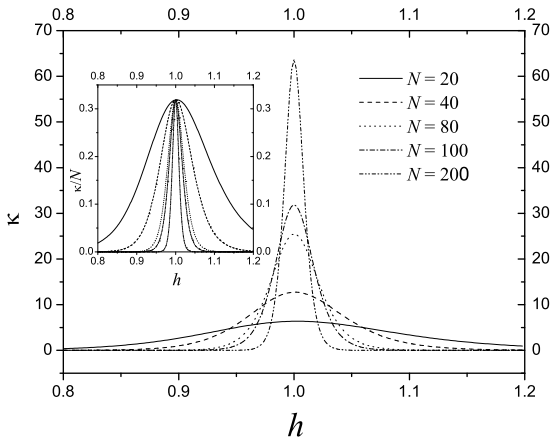}
\caption{(color online) Spin torsion modulus as a function of $h$ for
various system size. The inset denotes the normalized spin system modulus.}
\label{fig:spinmodulus}
\end{figure}

In this section, we are going to study the mechanical response of the spin
chain under a small rotation. In mechanics, a flexible elastic spring stores
mechanical energy when it is twisted. As long as the spring is not twisted
beyond its elastic limit, it obeys an angular form of Hooke's law%
\begin{equation}
\tau =-\kappa \phi ,
\end{equation}%
where $\tau $ is the torque exerted by the spring and $\kappa $ is a
constant called the spring's torsion elastic modulus. The potential energy
stored in the spring is defined as
\begin{equation}
U=\frac{1}{2}\kappa \phi ^{2}.
\end{equation}

For the present model, if we let the spin chain along the $z$-direction, the
rotation performed on the spin chain behaves like spin torsion. Then we can
introduce the spin torsion modulus as
\begin{equation}
\kappa =\left. \frac{d^{2}E_{0}}{d\phi ^{2}}\right\vert _{\phi =0}.
\label{eq:kappa}
\end{equation}%
For arbitray $\phi $, the energy spectra of the Hamiltonian is not the same
as the one-dimensional tranverse field Ising model. The boundary conditions
in Eq. (\ref{eq:Hamiltonianfermi}) are twisted boundary conditions.
Nevertheless, the Hamiltonian can be solved and ground-state energy can be
still calculated as%
\begin{equation}
E_{0}(\phi )=-\sum_{k}\sqrt{1-2h\cos (k-\phi )+h^{2}}.
\end{equation}
From Eq. (\ref{eq:kappa}), the spin torsion modulus of the ground state is

\begin{equation}
\kappa =\sum_{k}\frac{h^{2}\left[ 1+\cos ^{2}(k)\right] -h(1+h^{2})\cos (k)}{%
\left[ 1+h^{2}-2h\cos (k)\right] ^{3/2}}.
\end{equation}

Fig. \ref{fig:spinmodulus} shows the spin torsion modulus as a function of $%
h $ for various system sizes. From the figure, we can see that the
normalized modulus is non-zero only at the critical point. Away from the
critical point, it is almost zero. The reason is that in the non-critical
region, the systems is gapped, the ground state keeps unchanged under a
small rotation due to the gap protection. Hence only at the critical point,
the normalized modulus is nonzero. This observation can also be obtained
exactly. At the critical point%
\begin{equation}
\kappa =\sum_{k}\frac{1+\cos ^{2}(k)-2\cos (k)}{\left[ 2-2\cos (k)\right]
^{3/2}},
\end{equation}%
which at the infinite limit $\kappa /N=1/\pi $. While for $h\neq 1$, $\kappa
/N\rightarrow 0$.

The spin torsion modulus will become the spin stiffness if the $z$-component
of the Hamiltonian is conserved. This situation is happened at the isotropic
case of the one-dimensional XY model, whose Hamiltonian reads
\begin{equation}
H_{\text{XY}}=-\sum_{j=1}^{N}\left[ \frac{1}{2}\left( \sigma _{j}^{x}\sigma
_{j+1}^{x}+\sigma _{j}^{y}\sigma _{j+1}^{y}\right) +h\sigma _{j}^{z}\right] .
\end{equation}%
Under the rotation, the XY model can be transformed to a fermionic model
with twisted boundary conditions%
\begin{equation}
H_{\text{XY}}(\phi )=-\sum_{j=1}^{N}\left[ \sigma _{j}^{+}\sigma
_{j+1}^{-}e^{-i\phi }+\sigma _{j}^{-}\sigma _{j+1}^{+}e^{i\phi }+h\sigma
_{j}^{z}\right] .
\end{equation}%
The spin stiffness can be calculated as\cite{Kohn}
\begin{equation}
D_{c}=\left. \frac{1}{N}\frac{d^{2}E(\phi )}{d\phi ^{2}}\right\vert _{\phi
=0}.
\end{equation}%
For the XY model with twisted boundary conditions, the ground state energy
is $E=-\sum_{k}\left[ \cos (k-\phi )-h\right] $. The charge stiffness for $%
h=0$ is%
\begin{equation}
D_{c}=\frac{2}{\pi }.
\end{equation}

\section{Summary}

\label{sec:sum}

In summary, we studied the quantum phase transition occurring in the ground
state of a quantum spin chain spin with spiral orders in perspectives of
quantum information science and mechanical response. We show that the mode
of fidelity and its susceptibility can not only detect the quantum critical
point but also indicate which modes undergoes a significant change around
the transition point. Our results manifest that the mode fidelity and its
susceptibility are much singular that the global fidelity and its
susceptibility. From this point of view, we guess they are more suitable to
detect high-order transition. We study also the spin torsion modulus of the
model. The normalized modulus tends to zero in the non-critical region, only
at the critical point, it becomes a constant. Therefore, the spin torsion
modulus may help us to understand quantum critical phenomena from a
mechanical point of view.

\textit{Note added.} During finishing the work, we received a preprint from
P. D. Sacramento who used the mode fidelity and susceptibility to study a
superconducting model\cite{Sacramento}.

\section{Acknowledgement}

This work is supported by the Earmarked Grant Research from the Research
Grants Council of HKSAR, China (Project No. HKUST3/CRF/09).


\begin{thebibliography}{99}
\bibitem{Yoshimori} A. Yoshimori, J. Phys. Soc. Jpn. 14, 807 (1959).

\bibitem{Kaplan} T. A. Kaplan, Phys. Rev. 116, 888 (1959).

\bibitem{Freiser} M. J. Freiser, Phys. Rev. \textbf{123}, 2003 (1961).

\bibitem{Redner} S. Redner and H. E. Stanley,Phys. Rev. B 16, 4901 (1977).

\bibitem{Rosenfeld} E. V. Rosenfeld, N. V. Mushnikov, and V. V. Dyakin,
Physica Status Solidi (b) 246, 2187 (2009).

\bibitem{Kaplan2009} T. A. Kaplan, Phys. Rev. B \textbf{80}, 012407 (2009).

\bibitem{JCao2003} J. Cao, H. Q. Lin, K. J. Shi, and Y. Wang, Nuc. Phys. B
663, 487 (2003).

\bibitem{Nielsen1} M. A. Nilesen and I. L. Chuang, \textit{Quantum
Computation and Quantum Information} (Cambridge University Press, Cambridge,
England, 2000).

\bibitem{Sachdev} S. Sachdev, \textit{Quantum Phase Transitions}, (Cambridge
University Press, Cambridge, UK, 2000).

\bibitem{PPfeuty70} P. Pfeuty, Ann. Phys. \textbf{57}, 79 (1970).

\bibitem{RJElliott70} R. J. Elliott, P. Pfeuty, and C. Wood, Phys. Rev.
Lett. 25, 443 (1970).

\bibitem{Pzanardi2006} P. Zanardi and N. Paunkovic, Phys. Rev. E \textbf{74}%
, 031123 (2006).

\bibitem{HQZhou0701} H. Q. Zhou and J. P. Barjaktarevic, J. Phys. A: Math.
Theor. \textbf{41} 412001 (2008).

\bibitem{WLYou07} W. L. You, Y. W. Li, and S. J. Gu, Phys. Rev. E \textbf{76}%
, 022101 (2007).

\bibitem{SChen07} S. Chen, L. Wang, S. J. Gu, and Y. Wang, Phys. Rev. E
\textbf{76} 061108 (2007).

\bibitem{Venuti07} L. C. Venuti and P. Zanardi, Phys. Rev. Lett. \textbf{99}%
, 095701 (2007).

\bibitem{SJGU08} S. J. Gu, H. M. Kwok, W. Q. Ning and H. Q. Lin, Phys. Rev.
B \textbf{77}, 245109 (2008).

\bibitem{HQZhou2008} H. Q. Zhou, J. H. Zhao and B. Li, J. Phys. A: Math.
Theor. \textbf{41}, 492002 (2008).

\bibitem{SJGUreview} S. J. Gu, Int. J. Mod. Phys. B \textbf{24}, 4371(2010).

\bibitem{Kohn} Walter Kohn, Phys. Rev. \textbf{133}, A171 (1964).

\bibitem{Sacramento} P. D. Sacramento, N. Paunkovic, and V. R. Vieira,
arXiv:1107.5931.
\end{thebibliography}
\end{document}